# Modelling long-term trends in lunar exposure to the Earth's plasmasheet

Mike Hapgood[1]

[1]{STFC Rutherford Appleton Laboratory, Chilton, Didcot, Oxfordshire, OX11 OQX, UK}

Correspondence to: Mike Hapgood (M.Hapgood@rl.ac.uk)

## Abstract

This paper shows how the exposure of the Moon to the Earth's plasmasheet is subject to decadal variations due to lunar precession. The latter is a key property of the Moon's apparent orbit around the Earth – the nodes of that orbit precess around the ecliptic, completing one revolution every 18.6 years. This precession is responsible for a number of astronomical phenomena, e.g. the year to year drift of solar and lunar eclipse periods. It also controls the ecliptic latitude at which the Moon crosses the magnetotail and thus the number and duration of lunar encounters with the plasmasheet. This paper presents a detailed model of those encounters and applies it to the period 1960 to 2030. This shows that the total lunar exposure to the plasmasheet will vary from 10 hours per month at a minimum of the eighteen-year cycle rising to 40 hours per month at the maximum. These variations could have a profound impact on the accumulation of charge due plasmasheet electrons impacting the lunar surface. Thus we should expect the level of lunar surface charging to vary over the eighteen-year cycle. The literature contains reports that support this: several observations made during the cycle maximum of 1994-2000 are attributed to bombardment and charging of the lunar surface by plasmasheet electrons. Thus we conclude that lunar surface charging will vary markedly over an eighteen-year cycle driven by lunar precession. It is important to interpret lunar environment measurements in the context of this cycle and to allow for the cycle when designing equipment for deployment on the lunar surface. This is particularly important in respect of developing plans for robotic exploration on the lunar surface during the next cycle maximum of 2012-19.

Keywords: 2764 Plasma sheet; 2732 Magnetosphere interactions with satellites and rings; 7855 Spacecraft sheaths, wakes, charging

# 1 Introduction

The growing interest in lunar exploration necessitates a better understanding of the operating environment at the lunar surface. A key element in that environment is the charging of the lunar surface and the resulting electrodynamic environment on and just above the surface. There is growing observational evidence that the lunar surface can acquire negative potentials of several kilovolts (i.e. relative to the potential some Debye lengths above the surface) when exposed to strong fluxes of energetic electrons. Such potentials are a threat to operation of devices on the surface and may play an important role in dust transport. One important source of energetic electrons is Earth's plasmasheet, which the Moon sometimes encounters around the time of Full Moon. In this paper we adapt existing models of the plasmasheet and the Moon's orbit to explore how the plasmasheet moves with respect to the Moon. We then estimate the likelihood of Moon-plasmasheet encounters and show, for the first time, that this is strongly modulated over an 18 year cycle, driven by the precession of the Moon's orbit. This modulation is consistent with existing observations and is an important context for interpreting those observations. The specification of the lunar charging environment must take account of the 18-year cycle and in particular include awareness of high-risk periods when encounters are highly likely – the next being in 2012-19.

# 2 Context

The Moon has a complex orbit because it is subject to two strong gravitational forces – namely those of the Earth and the Sun. The latter is the stronger (by factor 2) so the Moon should be considered to orbit the Sun but in an orbit that is strongly perturbed by the Earth. As a result the Moon appears to orbit the Earth with a synodic period of 29.6 days, giving us the familiar monthly cycle of lunar phases. This cycle takes the Moon through the tail of Earth's magnetosphere for 4 or 5 days around the time of Full Moon. During this period it may encounter the tail plasmasheet and thus be exposed to the energetic electrons that are often found there.

The likelihood of such encounters is determined by the Moon's track across the tail (which is around Xgse = -60 Re) and the position of the plasmasheet in that region. As outlined in Figure 1 both phenomena are extended in Ygse direction, so what is critical in determining encounters is their relative location in the Zgse direction:

- The Moon's Z location varies over the course of a year due to the inclination of its orbit with respect to the plane of the ecliptic. The amplitude in Zgse is roughly iRm, where Rm is the distance to the Moon (60 Re) and i is the inclination of the Moon's orbit (5.15°). Thus the amplitude is $60 \times 5.15 \times \pi/180 = 5.5$ Re.

- The plasmasheet Z location also varies over the course of the year, this time due to the annual variation of ±23.4 degrees in dipole tilt (we neglect diurnal change here, but include it in the full model below). This annual variation arises because the near-Earth tail magnetic field is aligned with the internal dipole but the more distant tail is aligned with the solar wind (i.e. parallel to the plane of ecliptic). The transition between these two regimes lies around X=-10 Re. The result is that the distant plasmasheet behaves as if it lies parallel to the plane of the ecliptic but is attached to a hinge in the plane of geodipole equator at R=10 Re (see Figure 2). Thus the plasmasheet moves in Zgse with an annual amplitude of $10 \sin(23.4°) = 4.0$ Re.

These two phenomena have similar amplitudes, so the likelihood of Moon-plasmasheet encounters will be determined by their relative phases. The phase of the plasmasheet motion is synchronised with the seasons, since the annual variation of geodipole tilt is a simple consequence of the annual motion of the Earth's rotation axis with respect to the Sun. But the phase of the Moon's Z location varies steadily from year to year as a result of the precession of the Moon's orbit (which completes one revolution every 18.6 years).

Thus the encounter likelihood will vary from year to year as a result of this 18.6 year cycle. This cycle is therefore a critical context for interpreting observations of lunar surface charging and for assessing the risk that charging presents to lunar exploration activities. It is important to understand the cycle in detail and therefore we have undertaken a detailed study of Moon-plasmasheet encounters using up-to-date models of both phenomena.

## 3  Modelling

The encounters were modelled by tracking the movement of the Moon using a standard tool that gives its position in inertial coordinates. This position was then converted to an appropriate magnetospheric coordinate system so we could determine when it was in or close to the tail region. In these cases we then estimated the distance of the Moon from the magnetic neutral sheet that separates the two lobes of the tail. The Moon was considered to be in the plasmasheet if within a distance ΔZ of neutral sheet (i.e. we model the plasmasheet as a layer 2ΔZ thick centred on the neutral sheet). We take ΔZ=2 Re as representative of plasmasheet half-thickness.

These models were implemented in the IDL computer language and using a modular approach. This has allowed us to develop a number of applications that explore different aspects of Moon-plasmasheet encounters as shown in Figures 3 to 6 below. The approach also allowed us to utilise existing IDL modules for tasks such as calculating Moon position, carrying out magnetospheric coordinate transformations and handing time tags.

### 3.1 Neutral sheet

The neutral sheet is represented using the semi-empirical model of Tsyganenko et al. (1998) which was specifically developed to cover the tail out to X=-100 Re (and thus including the region crossed by the Moon at Xgse= -60 Re). This model exploited the then new tail magnetic field measurements available from Geotail. It also exploited the upstream IMF monitoring data from IMP-8 and Wind, so that the model could include the twisting of the neutral sheet due to IMF By. This model has recently been updated by Tsyganenko and Fairfield (2004), but the update focuses on Xgse ≥ -50 Re and thus does not cover the Moon's trajectory across the tail. We therefore use the 1998 model in this work.

A central feature of both models is a magnetospheric coordinate system termed Geocentric Solar Wind Magnetospheric (GSW) coordinates (see Appendix B for a discussion on coordinate system nomenclature). GSW is conceptually equivalent to the long-established Geocentric Magnetospheric (GSM) system, but differs in its treatment of magnetospheric aberration. This aberration arises because the Earth's orbit motion (30 km s$^{-1}$) is significant with respect to the solar wind speed. It causes tail magnetosphere to lag slightly behind the Earth in its motion around the Sun, so that the axis of the magnetosphere is rotated clockwise, as seen from north of the ecliptic plane, by an angle θ=30/Vsw - θ$_0$ where Vsw is the solar wind speed and θ$_0$ is GSE longitude from which the solar wind flows towards the Earth. For typical solar wind conditions (Vsw=400 km s$^{-1}$, θ$_0$=0), this angle is about 4 degrees. The GSW system is aligned with the aberrated magnetosphere such that the X axis lies along its principal axis (whereas Xgsm lies along the Earth-Sun line). The remaining elements of GSW are exactly equivalent to GSM, i.e. X is positive sunward, Z is the projection of the dipole axis on the plane perpendicular to X and is positive northward and Y, of course, completes a right-handed triad.

To support this system we wrote an IDL procedure, following the algorithms of Tsyganenko et al. (1998) to calculate the transformation matrix from GSM to GSW. This can then be used to convert Cartesian vectors between the two systems via matrix multiplication.

The Tsyganenko 1998 model specifies the GSW Z position of the centre of the neutral sheet as a function of GSW Y position plus the geodipole tilt and the Y component of the interplanetary magnetic field – see their equation 2. There is no explicit model dependence on GSW X position but there is an implicit dependence through the model parameters. The model provides different parameter sets for each of seven ranges covering the tail from X= -15 to -100 Re. The two furthermost bins have ranges -40 to -60 Re and -60 to -100 Re and thus cover the Moon's passage across the tail (the Moon is approximately 60 Re from Earth). The range boundary at -60 Re is a potential source of problems as it could cause discontinuities in our modelling. However, in practice, we have not encountered any problems.

The neutral sheet model has been implemented in our code using equation 2 of Tsyganenko et al. as discussed above. We derive the model's inputs as follows:

- The geodipole tilt for any date is available from our local coordinate transformation library (Hapgood, 1992).

- The IMF By component is generally taken as zero. This assumption is driven by the need to model plasmasheet encounters on many dates for which IMF data are not available. See Annex A for further discussion on By effects.

- The solar wind velocity (used to calculate transformations to GSW) is taken as 400 km s$^{-1}$ radially away from the Sun ($\theta_0$=0). The use of this fixed value is driven by the same constraint as for By. We need to model plasmasheet encounters on many dates for which velocity data are not available. In this case the GSW system is identical to the Aberrated Geocentric Solar Magnetospheric coordinate system.

### 3.2 Moon's orbit

There are a number of freely available codes for calculating the position of the Moon at any given time. We have used the moonpos.pro procedure that is available as part of the IDL

Astronomy User's Library, maintained by NASA Goddard Spaceflight Center (http://idlastro.gsfc.nasa.gov/). The user documentation for this routine reports that it gives the Moon's position with an accuracy better than 1 km, which is more than adequate for our present purpose. For example, if we wish to estimate lunar exposure to the plasmasheet with an accuracy of 10 minutes, it is sufficient to locate the Moon an accuracy equivalent to 10 minutes of lunar motion (600s @ 1 km s$^{-1}$ = 600 km).

The moonpos procedure allows us to calculate the geocentric inertial (GEI) coordinates (i.e. right ascension, declination and geocentric distance) of the Moon at any time; the angular coordinates are referenced to the mean epoch of date. Thus we can convert the Moon's position to standard magnetospheric coordinates, e.g. geocentric solar ecliptic (GSE), using our local Clustran coordinate transformation library based on the algorithms of Hapgood (1992, 1995).

The local implementation of moonpos was verified by finding the times of full moons (taken as GSE longitude = 180 degrees) in the years from 1970 to 2007. This showed good agreement ($\leq$ 2 minutes) when compared against selected values from published tables of Full Moon times (British Astronomical Association, 1970 & 2007).

## 4 Results

Using these models we can calculate the track of the Moon across the magnetotail at 30 minute intervals and estimate when it enters and exits the plasmasheet. To illustrate this figure 3 shows the calculated track for the period around the Full Moon in June 1999: The Moon has five encounters with the model plasmasheet – as shown by the track segments in red. There are two medium duration encounters (25.0 and 18.5 hours respectively) and three short encounter (5.5, 3.5 and 6 hours).

The duration of these encounters is an important factor in lunar charging. When exposed to significant fluxes of energetic electrons, dielectric materials can accumulate charge over many days with little loss due to internal conductivity – and thereby develop high negative potential. This effect is well-known from charging studies on spacecraft in geosynchronous orbit and similar considerations should apply to lunar rocks and regolith.

Thus we calculate the durations of the plasmasheet encounters for each passage of the Moon through the magnetotail and estimate the total exposure of the Moon to the plasmasheet during each passage (i.e. each lunation). We have calculated this value for every tail passage in the 71 years from 1960 to 2030. This gives a good mix of past events (so we can compare with information in the literature and in existing databases) and future events (so we can compare with plans for lunar exploration).

The results of these calculations are summarised in figure 4, which shows how the total exposure to the plasmasheet varies over the study period. The monthly exposure (red curve) shows a very marked modulation with period around 18 years arising from the 18.6 years precession of the Moon's orbit. On top of this long-term modulation there is a substantial short-term modulation, whose amplitude exhibits a double peak around the maximum of 18-year cycle. To better distinguish these two features we proceed as follows:

(a) to highlight the long-term modulation we smooth the monthly exposures with a 25-month running mean (blue curve). This shows that the long-term modulation has a single broad peak.

(b) to highlight the amplitude of the short-term modulation we find the maximum and minimum monthly values in each half-year period centred on a solstice (upper and low green curves). The upper curve shows a marked dip around the maximum of each 18-year cycle, but the lower curve shows no systematic trend. Thus we conclude that the short-term modulation has a double peak around each of the long-term maxima. This is reinforced by Figure 5, which shows the difference of the two green curves in Figure 4.

We can better resolve the short- and long-term variations by plotting the total exposure as a function of year and the sequential number of each lunation with the year (1 to 13) as shown in Figure 6 (in cases where there are only 12 lunations in a calendar year, we fill the 13th lunation with the exposure value from the 1st lunation of the following year). Figure 6 shows the 18 year modulation that was prominent in Figure 4 and again shows the double peak in the period of greatest exposure. For example, note the pairs of red regions around lunation 6 in 1976/1980 and 1995/1999.

The minimum of the 18 year cycle occurs when the ascending node of the Moon's orbit (i.e. the point at which it crosses the plane of the ecliptic northbound) is close to the First point of Aries (right ascension 0°). In this configuration the Moon will have its most northerly tail crossing in December and most southerly tail crossing in June. This is in anti-phase with the annual motion of the plasmasheet as shown in Figure 2. Thus the Moon will only occasionally encounter the plasmasheet. Nine years later the configuration is reversed: the ascending node of the Moon's orbit is close right ascension 180°, so its most northerly tail crossing is in June and most southerly in December. This is now in phase with the annual motion of the plasmasheet, so the Moon will frequently encounter the plasmasheet. This cycle is summarised in Table 1.

The double peak in the short-term modulation can now be understood as a consequence of slight differences between ranges of Moon and plasmasheet locations in the Z direction (perpendicular to the ecliptic). As discussed above the Moon's Z location varies by ±5.5 Re while that of the plasmasheet varies by ±4 Re. Thus the best phase match between lunar tail crossings and plasmasheet location will be offset slightly to either side of the Moon reaching its greatest Z variation.

## 5   Discussion

There are several published observations attributed to particle impact on the lunar surface during plasmasheet encounters. For example, Haleskas et al. (2005, 2007) have reported lunar surface charging up to several thousand volts. They observed several hundred cases of upward-flowing electron beams with the Electron Reflectometer instrument on Lunar Prospector in 1998-99. They attributed these beams to acceleration by a potential between the lunar surface and the spacecraft in orbit around the Moon and estimated that potential from the energy of the beam. They found the events were associated either with crossings of the plasmasheet or with solar energetic particle events (see Figure 3 of Halekas, 2007).

A very different approach was reported by Wilson et al. (2006). They used a ground-based wide-field camera to observe emissions from sodium in the lunar exosphere and assess how the exosphere responds to magnetospheric influences. They report observations during five lunar eclipses (to reduce scattered light from the Moon itself) and were able to detect the exosphere out to 10 to 20 lunar radii. They found a marked differentiation within their data – three cases (in 1996-97) showed significantly higher exospheric densities than the other two. The three high density cases were correlated with recent (< 20 hours) plasmasheet encounters by the Moon, while in the other two cases the Moon had not encountered the plasmasheet for more than 40 hours. Wilson et al. therefore attributed the enhanced sodium exosphere to the release of sodium from the lunar surface as a result of bombardment of that surface by energetic particles in the plasmasheet.

These observations all fall during a peak (years 1994-2000) of the eighteen-year cycle discussed in this paper. Thus the lunar exposure to the plasmasheet during these cases will be relatively long compared to other parts of the cycle. This adds weight to the association of these observations with plasmasheet particle impacts on the lunar surface; these observations were made at the best phase of the eighteen-year cycle.

In contrast the Apollo missions to the Moon fell during a minimum (years 1966-1972) of the eighteen-year cycle. Furthermore, the Apollo missions were timed so that the spacecraft were in lunar orbit and on the surface just after First Quarter lunar phase. This was selected to provide optimum conditions for viewing the landing sites (moderate shadows to reveal surface features) and good duration of daylight after landing. This selection also ensured that the Moon was on the dusk flank of the magnetosphere and thus was unlikely to encounter the plasmasheet. A detailed analysis suggests that most missions were outside the magnetosphere when at the Moon, but that Apollo 12 and 15 may have just crossed the dusk magnetopause a few hours before leaving lunar orbit. With this minor caveat it is clear that the Apollo missions were not directly exposed to plasmasheet charging of the lunar surface.

## 6   Conclusions

The model described here shows, for the first time, that the exposure of the Moon to the plasmasheet is strongly modulated over an eighteen-year cycle driven by the precession of the Moon's apparent orbit around the Earth. The typical plasmasheet exposure in each tail crossing may vary from 10 to 40 hours over the eighteen-year cycle. These changes in exposure time are significant for the accumulation of charge in the dielectric materials that form the lunar surface. Thus they are likely to modulate the electric fields that develop above the surface as a result of energetic electron impact on the lunar surface. There is observational evidence to support the existence of electron bombardment and charging during the cycle maximum in the mid to late 1990s.

The eighteen-year cycle provides a context for interpreting other observations of lunar charging. For example we are currently (2003-2010) in a minimum of the cycle. This minimum spans the recent operation of ESA's SMART-1 mission as well as that of upcoming lunar missions such as NASA's Lunar Reconnaissance Orbiter and India's Chandryaan. It will be interesting to search for evidence of plasmasheet charging in the data from those missions, but the absence of charging signatures will not be conclusive.

However, it is important to understand lunar surface charging as this could have a number of important practical effects for lunar exploration:

a.	Dust transport on the lunar surface. We know from Apollo that dust has the potential to interfere with operations on the surface. Thus it is important to understand its properties, including transport mechanisms. Charging is significant here as strong electric fields may be able to levitate dust and transport dust above the surface. The existence of dust high above the surface is suggested by observations of "horizon glow" and "streamers" on the lunar terminator at the time of the Apollo missions (Zook and McCoy, 1991).

b.	Equipment on the surface can accumulate charge. This may interfere with operations of that equipment, e.g. through direct accumulation in sensors or by accumulation and discharge in dielectric materials. If the discharge is abrupt (i.e. arcing), it can generate false electrical signals (leading to anomalous behaviour); in severe cases this can damage equipment. Charging effects can be mitigated by careful design, but it is very desirable to characterise the environment in order to inform efficient design.

c.	Vehicles landing on the surface may have a different potential to the surface so there is a risk of discharge on landing. This can be mitigated by design (aircraft on Earth have similar problems). But again it is very desirable to characterise the environment in order to inform that design.

Future work on plasmasheet exposure should include both data analysis and further modelling work. On data analysis side we should assess whether there are any other existing datasets that can be studied. For example, the instruments left by Apollo on and around the Moon continued to operate into the late 1970s, overlapping with the cycle maximum of 1975-1982. However, the publicly available Apollo data at NSSDC contains only solar wind plasma measurements. Further work is needed to see if other datasets exist and can be exploited. On the modelling side, we should explore how By affects lunar exposure to the plasmasheet as discussed in Annex A below. This will require a major improvement of the existing model so that we can exploit the good quality By measurements available from 1996 onwards. It also requires major work to include solar cycle modulation of By.

It is also important to extend this modelling to estimate the fluences (i.e. time-integrated flux) of electrons to which the Moon is exposed. This will require the development of a quantitative model of electron fluxes in the plasmasheet and its coupling to a model of plasmasheet location with respect to the Moon, such as that presented in this paper. We conclude that it is timely to develop models of the particle fluxes in the plasmasheet. Such models are an important tool for specifying the environment that will be experienced by future lunar explorers, both robotic and human. These models will also have application in providing better estimates of the noise in spacecraft sensors due to bremsstrahlung from plasmasheet electrons. Previous work on this problem, e.g. for ESA's Newton-XMM X-ray astronomy mission, could not consider plasmasheet motions (e.g. see technical report available on http://epubs.cclrc.ac.uk/work-details?w=36658).

**Appendix A: A preliminary assessment of By effects on plasmasheet exposure**

The Tsyganenko (1998) plasmasheet model includes a dependence on the By component of the interplanetary magnetic field. We have neglected that dependence in the present work because By values are not available for most past dates and all of the future dates. To assess the likely impact of this neglect we carried out a short and simple study using By values from NSSDC's OMNI database, which are provided as hourly averages. However, the results were inconclusive for a number of reasons:

1. Most importantly the OMNI database has limited time coverage except for two key periods: the International Magnetospheric Study (1979-82) and the current era (1995 to now) started by the International Solar-Terrestrial Physics programme. Thus we can only apply these data to limited parts of the eighteen-year cycle; we cannot make a satisfactory study of a whole cycle.

2. Secondly, the study of long-term trends in By needs to take account of the now well-established solar cycle variations in the interplanetary magnetic field (Hapgood et al., 1991). This complicates the analysis of the limited periods discussed above; we will need to consider how the solar-cycle interacts with the eighteen-year cycle.

3. Finally the use of hourly By data is unsatisfactory because the interplanetary magnetic field is often variable on much shorter timescales. This could cause the plasmasheet to change position abruptly from hour to hour and thus our model cannot accurately estimate lunar exposure. We will need to use By data with shorter time samples in order to track the plasmasheet motion in a smooth way.

For these reasons we conclude that a proper inclusion of By effects in this modelling will require significant extra work beyond the scope of the present paper. The assessment of solar-cycle effects is an interesting challenge, while the use of higher time resolution By values will require significant code development to ensure that the model can be run in a reasonable time.

**Appendix B. Co-ordinate system nomenclature.**

The literature contains a variety of names to describe the aberrated coordinate system used in the neutral sheet models discussed above. The name Geocentric Solar Wind Magnetospheric (GSWM) coordinates was introduced in the 1998 paper by Tsyganenko et al. But, in the later paper by Tsyganenko and Fairfield (2004), this was altered to Geocentric Solar Wind coordinates (GSW) for consistency with the earlier definition of that system by Hones et al. (1986).

**Acknowledgements**

The author thanks NASA GSFC for providing access to the IDL Astronomy Users' Library and thanks the software developers who contributed moonpos.pro, and its supporting routines, to this Library. He also acknowledges use of OMNI database available at the US National Space Science Data Center. He thanks the referees for their helpful and stimulating comments.

Table 1. The right ascension of the ascending node (RAAN) of Moon's orbit determines the right ascensions (RA) at which the Moon has its maximum northerly and southerly latitudes with respect to the ecliptic. These then set the months in which the Full Moon (and associated tail crossing) reach their greatest distances (north and south) from the ecliptic plane. To understand this table it may be helpful to recall the Full Moon is opposite the Sun in the sky, thus the RA of the Full Moon is 180° from that of the Sun. For example, the Full Moon is at RA=90° when the Sun is at RA=270°, which is in December.

| RAAN | RA max north | Full Moon at RA max north | RA max south | Full Moon at RA max south |
| --- | --- | --- | --- | --- |
| 0° | 90° | December | 270° | June |
| 90° | 180° | March | 0° | September |
| 180° | 270° | June | 90° | December |
| 270° | 0° | September | 180° | March |

**Figure captions**

Fig. 1. Lunar crossing of the magnetotail around the full moon on 24 December 2007. The track of the Moon is projected on to the GSE YZ plane with markers at showing the position at the start of each UTC day. The tail magnetopause is indicated by the dotted circle; it is offset in the positive Y direction to reflect the aberration of the magnetosphere due to a 400 km s$^{-1}$ solar wind. The red region indicates the model plasmasheet location at the mid-point of the crossing (see text for details). An animated version of this figure, covering the whole tail crossing, is available at (link TBD).

Fig 2. A noon-midnight cut through the magnetosphere around the June solstice. The black lines show the field topology derived from the Tsyganenko 2005 model. The red line shows how the neutral sheet location can be represented by a simple hinge model. One segment lies in the geodipole equatorial plane while the other lies parallel to the ecliptic; the two are linked by a hinge at geocentric distance of 10 Re.

Fig 3. Lunar crossing of the magnetotail around the full moon on 28 June 1999 using the same format as Fig. 1. The periods in the plasmasheet are indicated by red segments on the Moon's track.

Fig. 4. Predicted lunar exposure to the plasmasheet as a function of time over the period 1960 to 2030. The red curve shows the total exposure to the plasmasheet during each monthly crossing of the magnetotail. The blue curve shows the effect of smoothing the red curve with a 25-month running mean. The green curves show the maximum and minimum monthly exposures in half-yearly bins centred on the solstices.

Fig 5. The short-term modulation in predicted lunar exposure to the plasmasheet. The red curve shows the difference between the half-yearly maxima and minima in monthly lunar exposure (as derived from Figure 4). For reference the blue curve shows the long-term modulation in the form of the 25-month running mean exposure.

Fig 6. Seasonal variations in lunar exposure to the plasmasheet. This shows the same data as in Fig 4 but plotted as a function of both year and month (lunation).

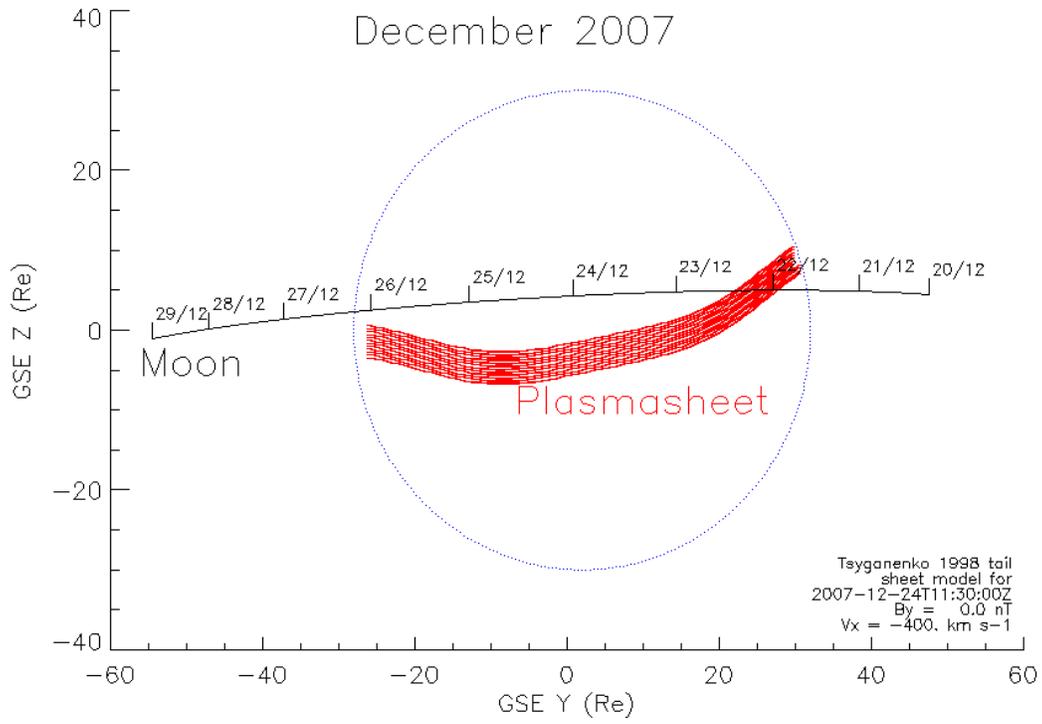

Figure 1

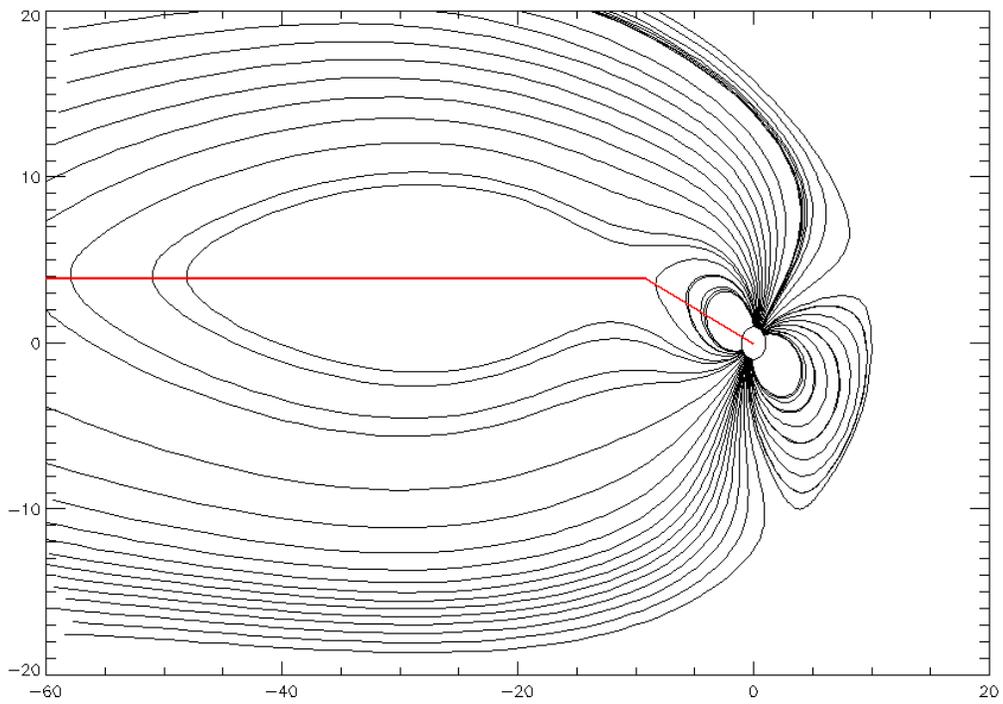

Figure 2

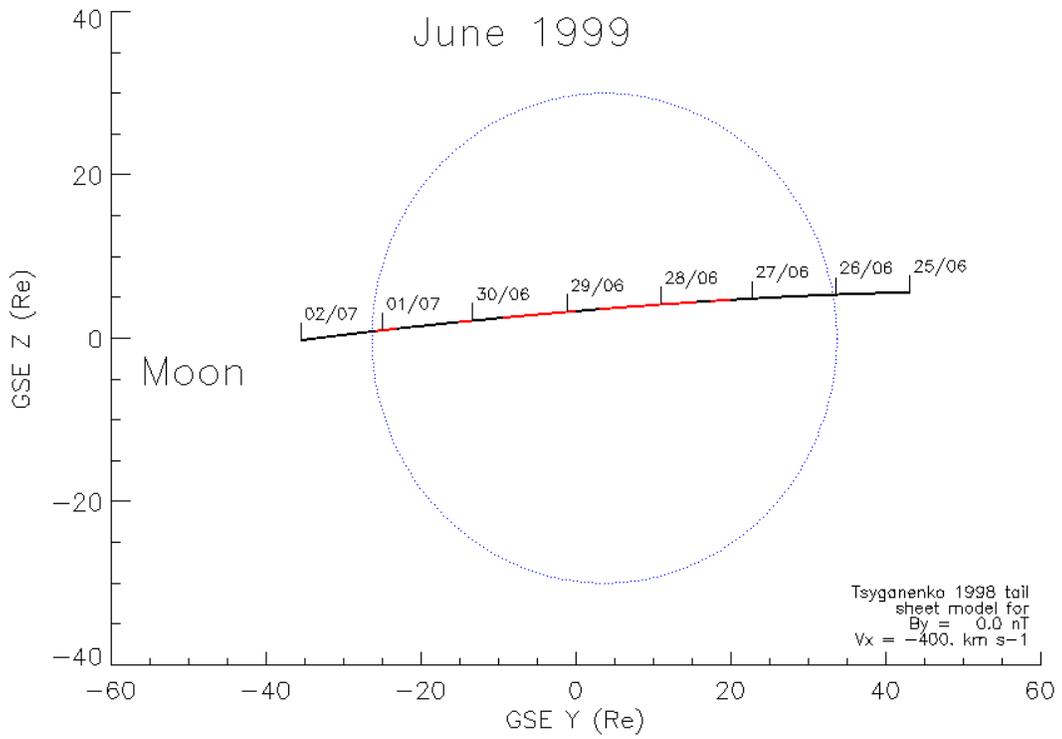

Figure 3

**Exposure per month**

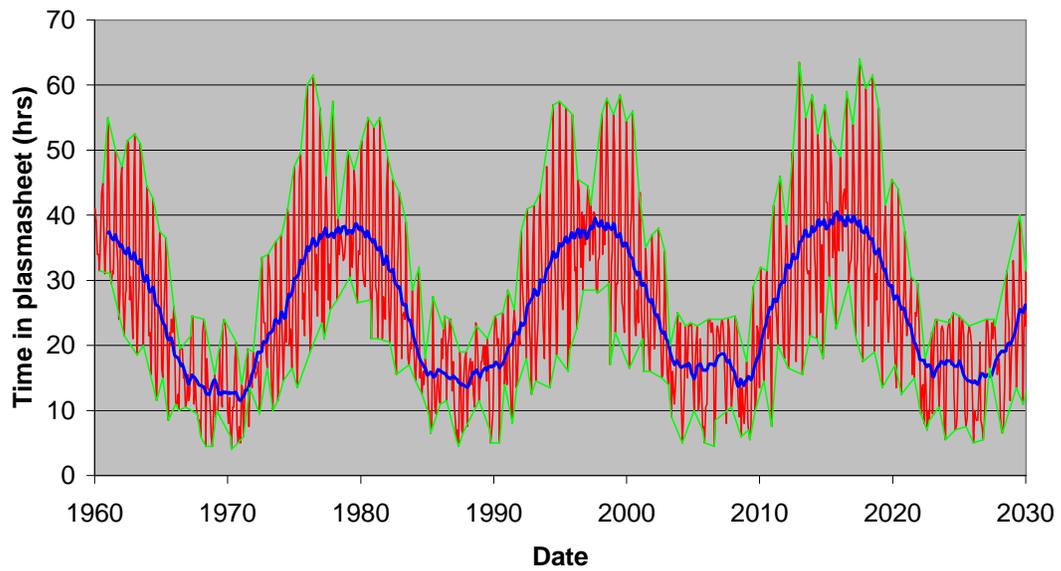

Figure 4

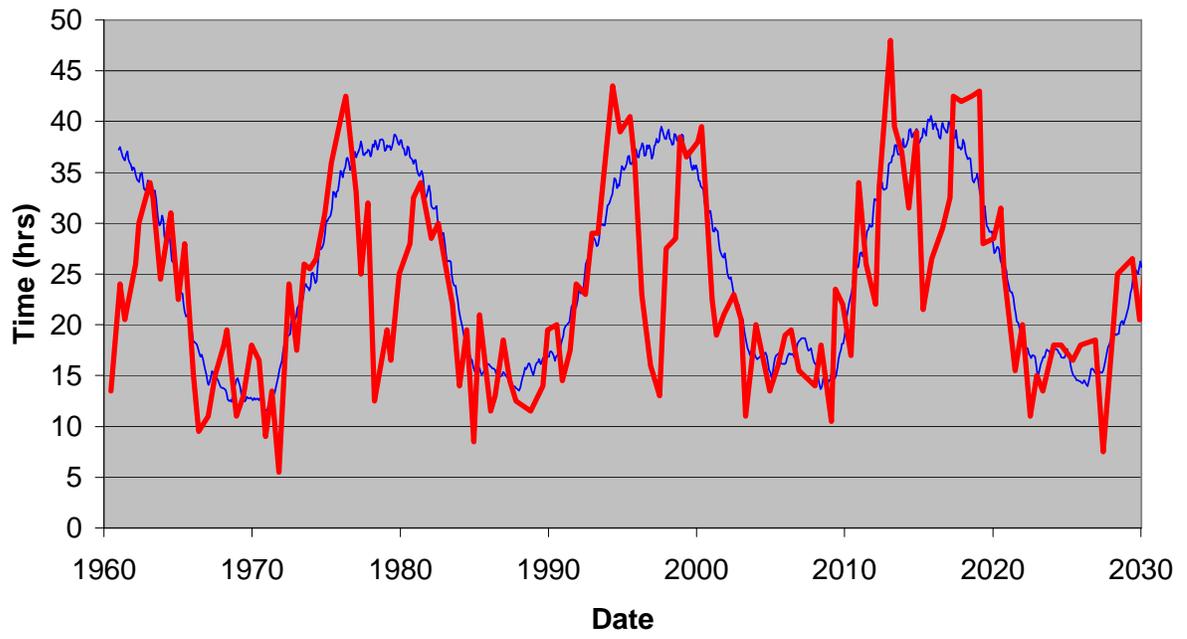

Figure 5

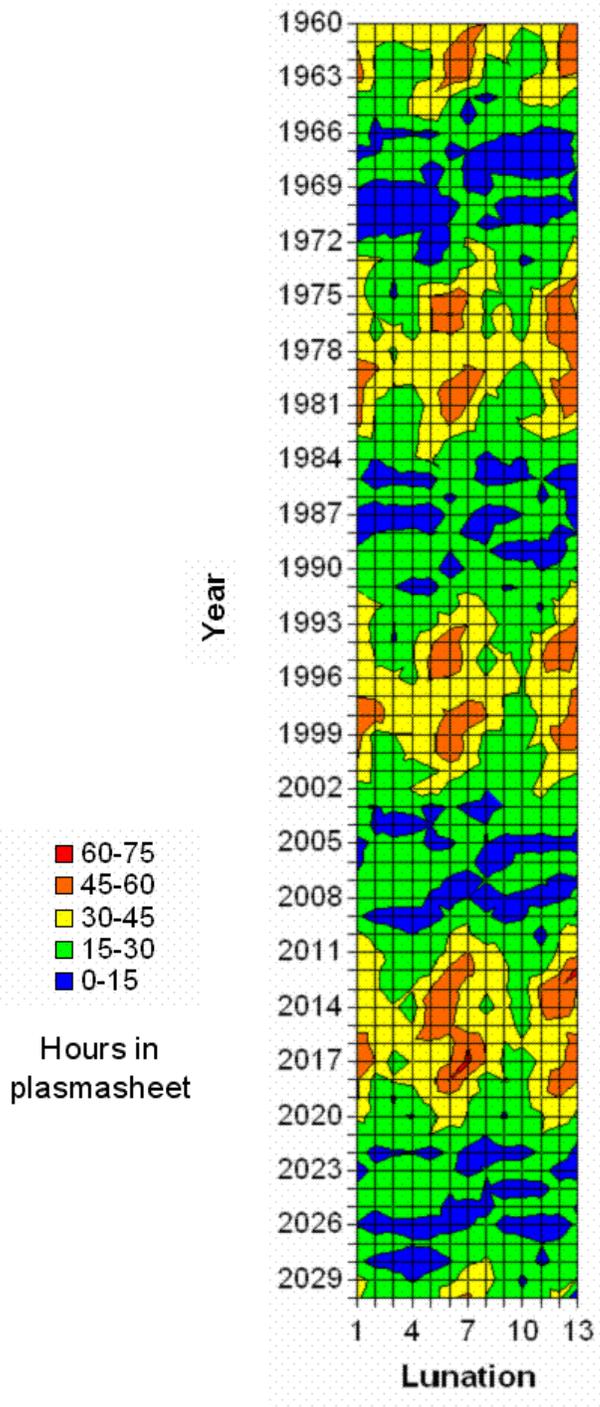

Figure 6